\documentclass[twocolumn,letterpaper,prl,superscriptaddress,showpacs]{revtex4-1}
\usepackage{color}
\usepackage{amsmath}
\usepackage{graphicx}
\usepackage{epsfig}
\usepackage{subfigure}
\usepackage{mathrsfs}

\begin{document}

\title{Calculation of Multipolar Exchange Interactions in Spin-Orbital Coupled Systems}
\author{Shu-Ting Pi}
\email{spi@ucdavis.edu}
\affiliation{Department of Physics, University of California, Davis, One Shields Avenue, Davis, California 95616 USA}

\author{Ravindra Nanguneri}
\affiliation{Department of Physics, University of California, Davis, One Shields Avenue, Davis, California 95616 USA}

\author{Sergey Savrasov}
\email{savrasov@physics.ucdavis.edu}
\affiliation{Department of Physics, University of California, Davis, One Shields Avenue, Davis, California 95616 USA}

\begin{abstract}
A new method of computing multipolar exchange interaction in spin-orbit
coupled systems is developed using multipolar tensor expansion of the
density matrix in LDA+U electronic structure calculation. Within mean-field
approximation, exchange constants can be mapped into a series of total
energy calculations by pair-flip technique. Application to Uranium dioxide
shows an antiferromagnetic superexchange coupling in dipoles but
ferromagnetic in quadrupoles which is very different from past studies.
Further calculation of spin-lattice interaction indicates it is of the same
order with superexchange and characterizes the overall behavior of
quadrupolar part as a competition between them.
\end{abstract}

\maketitle

Magnetic systems with strong spin-orbit coupling have been a theoretically
challenging problem for decades due to their complex magnetic behavior and
the lack of efficient computational techniques to solve model Hamiltonians
describing them. They not only have active orbital degrees of freedom, which
make these systems rich in magnetic properties, but they also possess a
large number of parameters in the form of corresponding inter-site exchange
interactions \cite%
{CS-1,CS-2,CS-3,CP-1,CP-2,CP-3,CAB-1,CAB-2,CAB-3,CAB-4,CAB-5,TR-1}. In 60s,
Schrieffer et. al. proposed a framework regarding the exchange interactions
mediated by RKKY mechanism in such systems \cite{CS-1,CS-2,CS-3}. Unlike
traditional spin $\frac{1}{2}$ problem where a simple Heisenberg model
describes the low-energy physics well \cite{SE-1}, the orbital degrees of
freedom introduce more complicated multipolar exchange couplings,
accompanied by large inter--site anisotropy, which makes the problem
computationally difficult \cite{TR-1}. In 80s, Cooper et. al. solved the
Coqblin--Schrieffer Hamiltonian for Cerium compounds and, in 90s, proposed a
scheme to compute the exchange constants via advanced electronic calculations \cite{CP-1,CP-2,CP-3,CAB-1,CAB-2,CAB-3,CAB-4,CAB-5}. Although their
works are in good agreement with experiments for selected simple materials,
an efficient and systematic method to calculate the exchange interaction is
still lacking.

In this work, we introduce a new method combined with electronic structure
calculations based on density functional theory (DFT) in its local density
approximation (LDA) or including the correction due to Hubbard U
via so--called LDA+U method \citep{LDA-1}, to compute the exchange
interactions of systems with strong spin-orbit coupling (SOC). It is based
on the theorem that multipolar tensor harmonics form a complete orthonormal
basis set with respect to the trace inner product. Applying this theorem to
the density matrix of the correlated magnetic orbital, well--defined scalar,
dipole, quadrupole, and higher multipoles can be extracted \cite{TR-1}. By
flipping a pair of tensor harmonics with respect to the ground--state
density matrix, we can find the exchange interaction by relating (or
mapping) it to the total energy cost of the tensor flip (which is obtained
by the LDA+U calculation).

To test our new method, we use Uranium Dioxide (UO$_{2}$) as a test
candidate due to the presence of dipolar and quadrupolar order in its ground
state. UO$_{2}$ has been one of the widely discussed actinide compounds due
to its applications in nuclear energy industry. It is a Mott insulator with
cubic structure and well--localized $5f^{2}$ electrons (Uranium
configuration U$^{4+}$ by naive charge counting). Below $T_{N}=30.8K$ it
undergoes a first--order magnetic and structural phase transition where a
noncollinear antiferromagnetic (AFM) phase with tranverse 3-$\mathbf{k}$
magnetic ordering accompanied by the cooperative Jahn--Teller distortion
occurs \cite{UE-2}. The two--electron ground state forms a $\Gamma _{5}$
triplet holding pseudospin $S=1$ rotation symmetry making it a good choice
to test our method, as it is a minimal challenge beyond $S=\frac{1}{2}$
Heisenberg model. Description of a $S=1$ exchange interaction requires the
existence of dipolar and quadrupolar moments, and it is commonly believed
that there are two major mechanisms to induce exchange coupling: 1)
superexchange (SE), and 2) spin--lattice interaction (SL). The former
contributes to both dipole and quadrupole and the latter contributes to
quadrupole only because of the symmetry of the distortion. The dominance of
SE or SL in affecting the quadrupole exchange remains a controversial issue
\cite{UE-1,UE-2,UE-3,UE-4}. Since our method is based on a static electronic
calculation, we do not explore dynamical effects in all their details.
Therefore, separate calculations using the coupled frozen--phonon and
frozen--magnon techniques were performed to extract the SL coupling
constants. Although we have chosen $UO_{2}$ as our test sample whose static
exchange interactions originate from superexchange mechanism, it should be
emphasized that our method should be able to work for any other types of
exchange processes.
\begin{figure}[tbp]
\centering
\includegraphics[width=1.0\columnwidth]{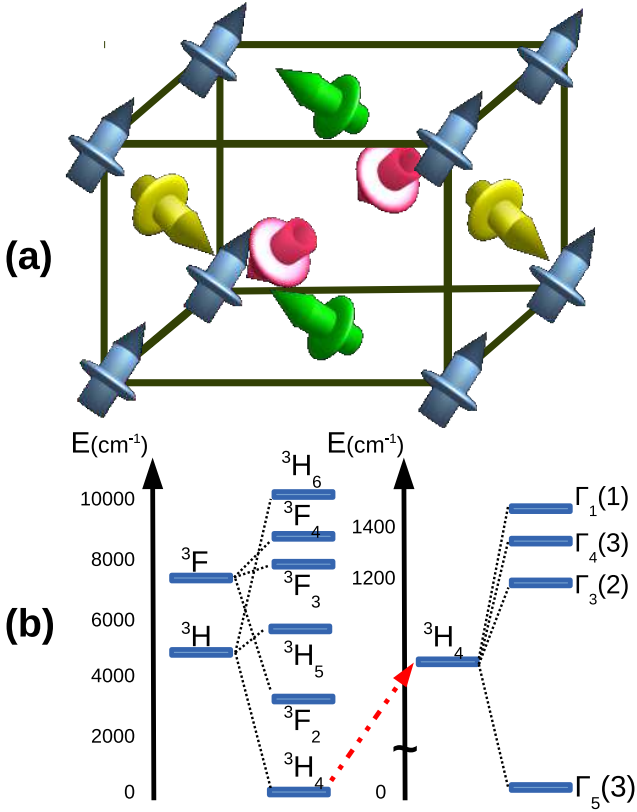}
\caption{(color) (a) Magnetic moments of dipole (arrows) and quadrupoles
(disk) in the 3-$\mathbf{k}$ structure. (b) The energy splitting of low
lying states of $UO_{2}$ \protect\cite{UE-2}. The $^{3}H$ and $^{3}F$ states
of free $U^{4+}$ ion is split into $^{3}H_{4}$ multiplets and other excited
states by spin-orbital coupling and further split into the $\Gamma _{5}$
triplet ground state by crystal fields. Inside the parentheses are their
degeneracy.}
\end{figure}


A non--hermitian unit spherical tensor operator is defined as: $%
Y_{K}^{Q}(J)=\sum_{MM^{\prime }}(-1)^{J-M}(2K+1)^{2}\times \left(
\begin{smallmatrix}
J & J & K \\
M^{^{\prime }} & -M & Q%
\end{smallmatrix}%
\right) |JM\rangle \langle JM^{^{\prime }}|$. We can further define
hermitian cubic tensor harmonics $T_{K}^{Q}(J)=\frac{1}{\sqrt{2}}%
[(-1)^{Q}Y_{K}^{Q}(J)+Y_{K}^{-Q}(J)]$ and $T_{K}^{-Q}(J)=\frac{i}{\sqrt{2}}%
[Y_{K}^{Q}(J)-(-1)^{Q}Y_{K}^{-Q}(J)]$ \cite{TR-1}. Since we only focus on $J=1$, the
label of $J$ will be omitted in the following. Based on the irreducible
representations of $J=1$ we can classify the cubic tensor harmonics as: $%
T^{s}$ for rank $0$ (scalar); $T^{x}$, $T^{y}$, $T^{z}$ for rank $1$
(dipole); $T^{xy}$, $T^{yz}$, $T^{zx}$, $T^{x^{2}-y^{2}}$, $T^{3z^{2}-r^{2}}$
for rank $2$ (quadrupole) \cite{RVH-1,UE-1,UE-3}. Since the $\Gamma _{5}$ triplet exhibit $S=1$
symmetry, it is convenient to denote them using the basis states: $|p\rangle
$, $p=+1,0,-1$. The ground state density matrix of an U ion can be expanded
by cubic tensor harmonics: $\rho _{i}=\sum_{m}\alpha _{i}^{m}T_{i}^{m}$,
where $i$ is site index, $m$ is the projection index for cubic harmonics,
and $\alpha _{i}^{m}=tr(\rho _{i}^{\dagger }T_{i}^{m})$ is the expansion
coefficient. Since the triplet degeneracy of $\Gamma _{5}$ is further split
below $T_{c}$, we can approximate the
ground state as $|GS\rangle =|-1\rangle $, the lowest energy state of an
isolated U-ion in the 3-$\mathbf{k}$ magnetic phase. 3-$\mathbf{k}$ ordering
requires the four $U$ sublattice moments all point in inequivalent $(1,1,1)$
directions, which means the $|-1\rangle $ states are defined in different
local coordinates for each $U$ sublattice \cite{UE-2}. Thus, we need to make a
rotation on each site to ensure everything is in a common global coordinate.
In the global system, the non--vanishing quadrupole components of the ground
state 3-$\mathbf{k}$ quadrupole order are $xy$, $yz$ and $zx$. Thus the
model Hamiltonian of nearest--neighbor exchange interaction between magnetic
$U$ atoms is assumed to be:
\begin{align}
h^{EX}& =h^{SE}+h^{SL} \\
&
=\sum_{m,i,j}C_{ij}^{mm}T_{i}^{m}T_{j}^{m}+%
\sum_{n,i,j}K_{ij}^{nn}T_{i}^{n}T_{j}^{n}  \notag \\
& m\in x,y,z,xy,yz,zx\quad ;\quad n\in xy,yz,zx,  \notag
\end{align}%
where ($i,j$) are nearest--neighbor site indexes and ($C_{ij}^{mm}$, $%
K_{ij}^{nn}$) are the exchange constants from SE and SL respectively.
Couplings between tensor operators with different symmetry indies are
prohibited by cubic symmetry. Since the coupling in $h^{SL}$ is a dynamical
effect, we will only focus on $h^{SE}$ part here and leave the $h^{SL}$ part
to a later discussion. The energy of $h^{SE}$ under mean field approximation
is $E_{0}=\langle h^{SE}\rangle \simeq 2\sum_{m,i,j}C_{ij}^{mm}\langle
T_{i}^{m}\rangle \langle T_{j}^{m}\rangle $. Suppose we make a
transformation of the tensor components of the density matrices on $U$ sublattices ($i$, $j$) in the
same unit cell, say in the components of $T_{i}^{xy}$ and $T_{j}^{xy}$: $%
\alpha _{i(j)}^{xy}\rightarrow \alpha _{i(j)}^{^{\prime }xy}$. If so, $%
\langle T_{i(j)}^{xy}\rangle \rightarrow \langle T_{i(j)}^{xy}\rangle
^{^{\prime }}=\langle T_{i(j)}^{xy}\rangle +\delta \langle
T_{i(j)}^{xy}\rangle $ with $\delta \langle T_{i(j)}^{xy}\rangle =(\alpha
_{i(j)}^{^{\prime }xy}-\alpha _{i(j)}^{xy})$. When we calculate the energy
difference between the transformed and original configurations, ($%
E^{^{\prime }}-E^{0})=(\langle h^{SE}\rangle ^{^{\prime }}-\langle
h^{SE}\rangle $), one can easily obtain a relation which is also true in
general for other exchange constants:
\begin{equation}
C_{ij}^{mn}=\frac{1}{8}\frac{\delta ^{2}E_{ij}^{mn}}{\delta \langle
T_{i}^{m}\rangle \delta \langle T_{j}^{n}\rangle },
\end{equation}%
where $\delta ^{2}E_{ij}^{mn}=(\delta E_{ij}^{mn}-\delta E_{i}^{m}-\delta
E_{j}^{n})$ is the interaction energy of the transformed pair; $\delta
E_{i}^{m}=(E_{i}^{m}-E_{0})$ is the energy cost from making a transformation
on the $T_{i}^{m}$ component, and, similarly, $\delta
E_{ij}^{mn}=(E_{ij}^{mn}-E_{0})$ is the energy cost from making
transformations on both $T_{i}^{m}$ and $T_{j}^{n}$ components. The
pre--factor $\frac{1}{8}=\frac{1}{4}\times \frac{1}{2}$ comes from the
correction for number of bonds between $U$--sublattice ($i$, $j$), the mean
field factor, as well as any geometric or trigonometric factor due to the
non-collinear order.

The basic idea of our method is to make the above transformations on the
density matrices of the correlated magnetic ions in the LDA+U calculation.
We then perform just one iteration in the self--consistent loop (to avoid
any change in the input density matrices) and compute the correlation energy
$\delta ^{2}E_{ij}^{mn}$ from the resulting band energies \cite{SE-1} as
prescribed by the Andersen force theorem \cite{AF-1}. Obviously, a single
exchange constant will need at least four values: no change ($E_{0}$),
single--site change ($E_{i}$,$E_{j}$) and double--site changes ($E_{ij}$).
The choice of the transformation has to preserve the symmetry of the crystal
field, the charge density, and the magnitude of magnetic moment to prevent
any unwanted energy cost. A reasonable choice is to ``flip" the orientation
of magnetic moment by adding a minus sign on the expansion coefficient of
the corresponding tensor component. When this is done, $\delta \langle
T_{i}^{m}\rangle $ is always $-2\langle T_{i}^{m}\rangle $, which is
equivalent to making a $\pi $ rotation on ($x$, $y$, $z$) components of the
dipole and a $\pi /2$ rotation on ($xy$, $yz$, $zx$) components of the
quadrupole.

To generate density matrices that are compatible with the single--particle
based LDA+U calculation, we introduce the reduced density matrix (RDM) as a
useful single--particle approximation to the $\Gamma _{5}$ states. \cite{RDM-1} We assume
that the multipolar exchange Hamiltonian in SOC f-orbital space is built by
replacing all tensor operators, density matrices, and mean values in $S=1$
space to their corresponding single--particle RDM: $\langle T_{i}^{m}\rangle
\rightarrow \langle \mathscr{T}_{i}^{m}\rangle $, $\langle \rho _{i}\rangle
\rightarrow \langle \mathscr{D}_{i}\rangle $. The single--particle exchange
Hamiltonian shares the same exchange constants as the $S=1$ two--particle
version. Two things to notice here are: 1) the RDM exhibits $J=\frac{5}{2}%
\oplus \frac{7}{2}$ symmetry instead of $S=1$ and this means the rotation
from local coordinates to the global coordinates has to be made in $S=1$
space, else the pseudospin quasi--particle description will be violated; 2)
the RDM replacement will rescale the length of an operator, i.e. $tr(%
\mathscr{T}^{\dagger }\mathscr{T})\neq tr(T^{\dagger }T)$. Therefore, $%
\langle \mathscr{T}_{i}^{m}\rangle =tr(\mathscr{D}^{\dagger }%
\mathscr{{T}^{m}_{i}})$ is different from $\langle T_{i}^{m}\rangle =tr(\rho
^{\dagger }T_{i}^{m})$. So one has to be cautious when using eq.(2).

\begin{table}[tbp]
\caption{Comparison between our calculated exchange interaction parameters
using the LDA+U method with $U=4.0$ eV and $J=0.7$ eV and the existing
experimental fits. $C_{0}^{d}$, $C_{0}^{q}$, $K_{0}^{q}$ are in units of
meV, others are dimensionless. Because all the works use different models
to simulate the SL part, there is no appropriate values for them (labeled
by *). Ref.\protect\cite{UE-4} obtained SL via a fully dynamic calculation
with long-range and frequency dependence. Note also that Ref.%
\protect\cite{UE-1} assumes the quadrupole coupling only comes from SL
with real space exchange constant of the 3-$k$ symmetric form: $%
K_{ij}^{\Gamma }=K_{0}e^{i\mathbf{q}_{\Gamma }(\mathbf{R}_{i}-\mathbf{R}%
_{j})}$. Ref.\protect\cite{UE-2} only calculates SE part. Their parameters
were obtained via the integrals of Coulomb interaction directly
and has no a simple anisotropy form.}%
\begin{ruledtabular}
\begin{tabular}{ccccccc}
Ref. &  $C_{0}^{d}$  & $\chi^{d}_{c}$ & $C^{q}_{0}$ & $\chi^{q}_{c}$ &
$K^{q}_{0}$ & $\chi^{q}_{K}$
\\ \hline
 ours & 1.70 & 0.3 & -3.10 & 0.90 & 2.6 & 1.18 \\
 \cite{UE-4} & 3.1 & 0.25 & 1.9 & 0.25 & $\ast$ & $\ast$  \\
 \cite{UE-1} & 1.25 & 0.8 & 0 & 0 & 0.33 & $\ast$ \\
 \cite{UE-2} & $\sim 1$ & $\ast$ & $\sim 0.1$ & $\ast$ & $\times$ & $\times$
\end{tabular}
\end{ruledtabular}
\end{table}

The coupling constants can be simplified by symmetry to the form: $%
C_{i,j}^{m,n}=C^{m,n}(\mathbf{R})=C_{0}^{d/q}[1-2(1-\chi _{c}^{d/q})\tau
_{m}\tau _{n}]\delta _{m,n}$, where $d/q$ means dipole or quadrupole and $%
\tau =\mathbf{R}/R$ is the direction vector between $(i,j)$. These constants
are shown in TABLE I, where the isotropic and anisotropic parts are
described by $C_{0}^{d/q}$ and $\chi _{c}^{d/q}$ respectively \cite{UE-1}.
With the comparison to other studies, the dipole part is similar, but the
quadrupole part gives the opposite result from past calculations obtained by
best fit with experiment \cite{UE-3,UE-4}. Not only the anisotropy effect is
much smaller, but the sign is also different which means the quadrupoles
tend to be ferromagnetic. It also means that the SL effects must be as
important as SE and their combination makes the whole system
antiferromagnetic.

\begin{figure}[tbp]
\centering
\includegraphics[width=1.0\columnwidth]{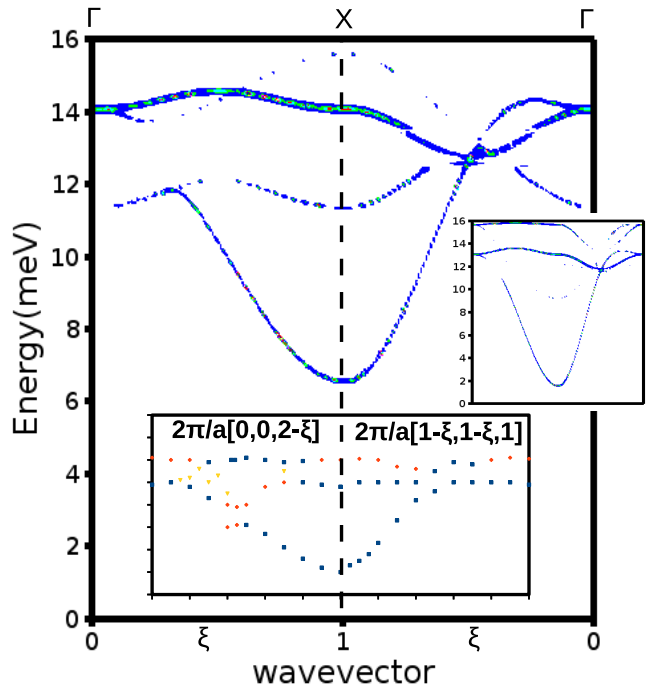}
\caption{(color) Magnetic excitation of $UO_{2}$ along two symmetry
directions calculated by scanning the colormap of the real part two-ion
susceptibility of our model Hamiltonian \protect\cite{UE-1} with parameters
shown in TABLE I. Right inset: The same calculation made by requiring the
overall quadrupole coupling to have 3-$\mathbf{k}$ symmetry: $K_{ij}^{\Gamma
}=K_{0}e^{i\mathbf{q}_{\Gamma }(\mathbf{R}_{i}-\mathbf{R}_{j})}$ with $%
K_{0}=0.5$ meV \cite{UE-1}. Anisotropy gap is greatly reduced. Bottom inset: data from
inelastic neutron scattering experiments plotted in the same $x-y$ scale%
\protect\cite{SW-1}. Triangles (yellow) are measured in a direction
differing by a reciprocal lattice vector\protect\cite{SW-2}. Rhombus
(orange) are weaker cross-section.}
\end{figure}


To explain the behavior of the quadrupolar part, we need to include the
effect of dynamic contribution from SL. The coupling between spins and
optical phonons can be written as:
\begin{equation*}
H_{SL}=\sum_{\mathbf{q},n,j}V^{n}(\mathbf{q},j)T^{n}(\mathbf{q})u(\mathbf{q}%
,j),
\end{equation*}%
where $T^{n}(\mathbf{q})=\sum_{\mathbf{R}}T^{n}(\mathbf{R})e^{i\mathbf{q}%
\cdot \mathbf{R}}$, $u(\mathbf{q},j)=[a^{\dagger }(-\mathbf{q},j)+a(\mathbf{q%
},j)]$ and $a^{\dagger }(\mathbf{q},j)$ is the creation operator of a phonon
with wavevector $\mathbf{q}$ in mode $j$. Using the virtual phonon
description, the SL exchange constant of $h^{SL}$ can be approximated as:
\begin{equation*}
K^{n,n}(\mathbf{q})\simeq \sum_{j}\frac{|V^{n}(\mathbf{q},j)|^{2}}{h\omega (%
\mathbf{q},j)}-\varepsilon _{0},
\end{equation*}%
where $\omega (\mathbf{q},j)$ is the phonon frequency and $\varepsilon _{0}$
is the onsite exchange energy which should be subtracted \cite{UE-1}. The
variables $u(\mathbf{q},j)$ and $\omega (\mathbf{q},j)$ have been calculated
in one of our earlier works \cite{ABP-1} and can be fitted to the entire
Brillouin Zone using a simple rigid--ion model \cite{RI-1,RI-2}. If we
further assume the quadrupoles only couple to $t_{2g}^{a}$ and $t_{2g}^{b}$
quadrupolar distortions of the O--cage around each U-ion, the coupling
constants are assumed to have the form: $V^{n}(\mathbf{q},j)=\gamma _{a}\psi
_{a}^{n}(\mathbf{q},j)+\gamma _{b}\psi _{b}^{n}(\mathbf{q},j)$, where $%
\gamma _{a/b}$ are the parameters to be determined, $\psi _{a/b}^{n}(\mathbf{%
q},j)$ are the inner product (projection) between the phonon distortion $u(%
\mathbf{q},j)$ and $t_{2g}^{a/b}$ distortion, and $u(\mathbf{q},j)$ can be
regarded as the distortion due to a phonon mode \cite{T2G-1}. We estimate
the parameters $\gamma _{a/b}$ by using a coupled frozen-phonon and
frozen--magnon technique: 1) Make a $t_{2g}^{a/b}$ distortion of the O-cage
around an U-ion; 2) Flip a particular tensor component of the single-ion RDM
on a particular site; 3) Calculate the correlation energies: $\delta
^{2}E_{a/b}^{mn}=[\delta E_{a/b}^{mn}-\delta E_{a/b}^{0n}-\delta E^{m0}]$,
where the first superscript is the symmetry index of the quadrupole and the
latter index is of $t_{2g}^{a/b}$. So $\delta ^{2}E_{a/b}^{mn}$ is the extra
energy of making ``flip+frozen phonon distortion" simultaneously compared to
the energies of individual ``flip" plus individual ``frozen phonon
distortion"; 4) Then the parameters are roughly: $\gamma _{a}\sim \delta
^{2}E_{a}^{mn}/\sqrt{2}\langle T^{m}\rangle \psi _{a}^{n}$ and $\gamma
_{b}\sim \delta ^{2}E_{b}^{mn}/\langle T^{m}\rangle \psi _{b}^{n}$ . There
is an factor $\sqrt{2}$ in $\gamma _{a}$ because when we make the same
displacement of each coordinate component, the length of the total
displacement is $\sqrt{2}$ larger than $t_{2g}^{b}$. By assuming the unit of phonon vibration
about $0.014\mathring{A}$ (as is the static Jahn-Teller distortion \cite%
{UE-2}) and making a $t_{2g}$ distortion $3\%$ of the lattice constant, we have: $\gamma _{a}=34meV$ and $\gamma _{b}=48meV$. We can access
nearest neighbor constants by calculating $K^{n,n}(\mathbf{q},j)$ at $%
\mathbf{q}=[0,0,0]$ and $\mathbf{q}=\frac{2\pi }{a}[1,0,0]$, and by a
subsequent fit to a cosine function with the onsite exchange energy assumed
to be the average of the curve \cite{UE-1}. We then have: $%
K_{i,j}^{m,n}=K^{m,n}(\mathbf{R})=K_{0}^{q}[1-2(1-\chi _{k}^{q})\tau
_{m}\tau _{n}]\delta _{m,n}$ with $K_{0}^{q}=2.6$ meV and $\chi
_{k}^{q}=1.18 $.

Combined with the superexchange contribution and using the Green's function
method with random phase approximation \cite{UE-1}, we calculate the
magnetic excitation spectrum that is shown in FIG. 2. We find that the
values and the characteristics of our results are basically in agreement
with experiment. The major difference is the disappearance of anti-crossing
at a few $\mathbf{q}$-points and much larger ansiotropy (gap) at $X$-point.
The disappearance of the anti-crossing is reasonable because it comes from
the coupling between magnon and phonon branches. As for the overestimated
anisotropy at $X$-point, it is believed to come from the oversimplified SL
model in our calculation. We have plotted the spin/quadrupolar wave spectrum
by enforcing the overall quadrupole coupling to have 3-$k$ symmetry as Ref.%
\cite{UE-4} with the parameter $K_{0}=0.5$ meV (which is almost the same value
as our isotropic part) and it gives a much smaller gap which fits the
experiments well (see FIG. 2). It demonstrates that a SL model which makes
the whole quadrupole coupling to have 3-$k$ symmetry will be helpful in
fitting the experiment but, in this case, the simple form of our model is
also lost.


In conclusion, we have developed a new and efficient method for computing
the exchange interactions in systems with strong spin-orbit coupling. With
its application to $UO_{2}$ , the superexchange mechanism is found to have
very interesting ferromagnetic quadrupolar coupling which has not been
previously reported. We also performed estimates of the spin--lattice
coupling via a similar technique and the overall behavior is accounted for
by combining both effects. An accurate description of the spin--lattice
interaction is still an issue and will be a subject for future work.

We are grateful to X. Wan and R. Dong for their helpful discussions. This work was supported by US DOE Nuclear Energy University Program under Contract No. 00088708.


\end{document}